# The first-passage area for drifted Brownian motion and the moments of the Airy distribution

Michael J Kearney

*Faculty of Engineering and Physical Sciences, University of Surrey,*
*Guildford, Surrey, GU2 7XH, United Kingdom*

Satya N Majumdar

*Laboratoire de Physique Théorique et Modèles Statistique, Université Paris-Sud.*
*Bât. 100 91405, Orsay Cedex, France*

Richard J Martin

*Quantitative Credit Strategy Group, Credit Suisse, One Cabot Square,*
*London, E14 4QJ, United Kingdom*

*Abstract*

An exact expression for the distribution of the area swept out by a drifted Brownian motion till its first-passage time is derived. A study of the asymptotic behaviour confirms earlier conjectures and clarifies their range of validity. The analysis leads to a simple closed-form solution for the moments of the Airy distribution.

The Airy distribution (not to be confused with the Airy function) occurs quite naturally in a remarkably diverse range of problems; see [1-3] and references therein. Examples arise when considering, e.g., the cost function for algorithms used for data storage [4], the internal path length of rooted trees [5], the nature of fluctuations in inventory processes [6], the area enclosed by planar random loops [7], the maximal relative height for fluctuating interfaces [8], solid-on-solid models [9] and Gaussian signals with $1/f^\alpha$ spectra [10], and the avalanche size in directed sandpile models [11]. In perhaps its most natural setting the Airy distribution characterises the area swept out by a Brownian excursion, i.e. the area under the Wiener process $W(t)$ conditioned such that $W(0) = 0$, $W(1) = 0$ with $W(t) > 0$ for $0 < t < 1$. See [2] for a derivation from a path integral perspective, and [12] for an excellent overview of the wider class of Brownian area problems, many of which share similar features. It is known that the area moments of a Brownian excursion and hence the moments of the Airy distribution are given by [13,14];

$$M_n = \frac{4\sqrt{\pi} n!}{\Gamma(\frac{3n-1}{2}) 2^{n/2}} K_n \tag{1}$$

with $K_0 = -1/2$, $K_1 = 1/8$ whilst for $n \geq 2$ one has the quadratic recursion

$$K_n = \left(\frac{3n-4}{4}\right) K_{n-1} + \sum_{j=1}^{n-1} K_j K_{n-j}. \tag{2}$$

Slightly different definitions of the Airy distribution exist in the literature [12] which are equivalent up to a scaling of the area variable, e.g., in [1] the moments $\mu_n = (\sqrt{8})^n M_n = 2\sqrt{\pi}\, \Omega_n / \Gamma(\frac{3n-1}{2})$, where $\Omega_n = K_n 2^{n+1} n!$ are the so-called Airy constants. In all cases, however, the recursion (2) plays a fundamental role, and it is this object we focus on. It is easy to iterate (2) for small values of $n$; however, it has been a long-standing challenge to identify a *simple* closed-form solution for $K_n$ which is valid for general $n$. Below we will show that for all $n \geq 1$

$$K_n = \frac{3}{4\pi^2} \int_0^\infty \frac{z^{3(n-1)/2}}{\text{Ai}^2(z) + \text{Bi}^2(z)} dz \qquad (3)$$

where $\text{Ai}(z)$ and $\text{Bi}(z)$ are the standard, linearly independent solutions of Airy's differential equation $F''(z) - zF(z) = 0$.

To derive this rather elegant result we will first solve another problem, similar to but distinct from the Brownian excursion problem described above, building on the work presented in [15]. This ancillary problem finds application in determining the limiting behaviour of various models associated with discrete time queues [16], cellular automaton traffic jam models [17] etc. To set the scene, consider a stochastic process, $y(t)$, which evolves via the Langevin equation

$$\frac{dy(t)}{dt} = -\eta + \xi(t) \qquad (4)$$

where $\xi(t)$ is a zero mean noise source with correlator $\langle \xi(t)\xi(t') \rangle = \delta(t-t')$ such that $\xi(t)dt = dW(t)$. Thus $y(t)$ is a *drifted* Brownian motion. Suppose the motion starts at $y(t=0) = x > 0$ with drift parameter $\eta > 0$, and let $t_f$ denote the time at which the process first crosses $y = 0$. Then the variable $A = \int_0^{t_f} y(t')dt'$ defines the area swept out by the process till this first-passage time. One would like to know the probability density, $P(A,x)$, of this area variable. This was considered in [15] where it was shown, using a backward Fokker-Planck technique, that the Laplace transform of the probability density $\tilde{P}(s,x) \equiv \int_0^\infty P(A,x)e^{-sA}dA$ satisfies (for convenience and clarity we work hereafter in units where $\eta = 1$)

$$\frac{1}{2}\frac{\partial^2 \tilde{P}(s,x)}{\partial x^2} - \frac{\partial \tilde{P}(s,x)}{\partial x} - sx\tilde{P}(s,x) = 0 \qquad (5)$$

subject to the boundary conditions $\tilde{P}(s, x=0) = 1$ and $\tilde{P}(s, x \to \infty) = 0$. The solution is given by

$$\tilde{P}(s,x) = e^x \frac{\text{Ai}(2^{1/3} s^{1/3} x + 2^{-2/3} s^{-2/3})}{\text{Ai}(2^{-2/3} s^{-2/3})} \ . \tag{6}$$

Using this result, various asymptotic results were presented in [15] relating to the moments $\langle A^n \rangle$ and the tail of the probability density $P(A,x)$ as $A \to \infty$ when $x$ is small. It was left as a challenge to formally invert (6) to derive an explicit expression for $P(A,x)$. We take up this challenge here.

Thus we seek to evaluate the contour integral

$$P(A,x) = \frac{1}{2\pi i} \int_{b-i\infty}^{b+i\infty} \tilde{P}(s,x) e^{sA} ds \tag{7}$$

with $\tilde{P}(s,x)$ given by (6). The approach taken is similar to that discussed in [18] in the context of considering the integral of the absolute value of a Brownian bridge. We first wish to modify the contour by closing it to the left. The continuation to the left hand side of the complex $s$-plane can be accomplished by providing a cut on the negative real axis. Thus one can represent $s = re^{i\varphi}$ where $|\arg s| < \pi$ or $-\pi < \varphi < \pi$. Fractional powers $s^\mu$ with $|\mu| < 1$ are to be interpreted as $s^\mu = r^\mu e^{i\mu\varphi}$ so that $|\arg(s^\mu)| = |\mu\varphi| < \pi$. It then follows that $\tilde{P}(s,x)$ is an analytic function on the cut-plane since the zeros of $\text{Ai}(z)$ are restricted to the negative real axis [1]. Now, if one considers that $\text{Ai}(z)$ vanishes exponentially fast as $z \to \infty$ [19]

$$\text{Ai}(z) \sim \frac{1}{2} \pi^{-1/2} z^{-1/4} e^{-\frac{2}{3} z^{3/2}} ; \quad |\arg(z)| < \pi \tag{8}$$

it follows that the Bromwich contour in (7) may be deformed to a Hankel-type contour around the branch cut, starting at $-\infty$, winding round the origin in an anti-clockwise fashion and ending up at $-\infty$, i.e.,

$$P(A,x) = \frac{1}{2\pi i} \int_{-\infty}^{(0^+)} \tilde{P}(s,x) e^{sA} ds. \quad (9)$$

By considering the contributions above and below the branch cut one then finds that

$$P(A,x) = \frac{e^x}{2\pi i} \int_0^\infty \left[ \frac{\mathrm{Ai}(2^{-2/3} e^{i2\pi/3} r^{-2/3}(1+e^{-i\pi} 2xr))}{\mathrm{Ai}(2^{-2/3} e^{i2\pi/3} r^{-2/3})} \right. \\ \left. - \frac{\mathrm{Ai}(2^{-2/3} e^{-i2\pi/3} r^{-2/3}(1+e^{i\pi} 2xr))}{\mathrm{Ai}(2^{-2/3} e^{-i2\pi/3} r^{-2/3})} \right] e^{-rA} dr. \quad (10)$$

This appears to be no more tractable than the original expression; however, a key simplification comes from exploiting the identity [19]

$$\mathrm{Ai}(ze^{\pm 2\pi i/3}) = \frac{1}{2} e^{\pm \pi i/3} \left[ \mathrm{Ai}(z) \mp i\,\mathrm{Bi}(z) \right] \quad (11)$$

whereupon (noting that $e^{\pm i\pi} = -1$) one can recast (10) in the form

$$P(A,x) = \frac{e^x}{\pi} \int_0^\infty \frac{\mathrm{Ai}(\tau_r - x\tau_r^{-1/2})\mathrm{Bi}(\tau_r) - \mathrm{Bi}(\tau_r - x\tau_r^{-1/2})\mathrm{Ai}(\tau_r)}{\mathrm{Ai}^2(\tau_r) + \mathrm{Bi}^2(\tau_r)} e^{-Ar} dr \quad (12)$$

with $\tau_r \equiv (2r)^{-2/3}$. This is an exact expression for the first-passage area probability density for drifted Brownian motion in terms of the integral of a *real* function, valid for all $A > 0$ and $x > 0$. It is, of course, not the same as the Airy distribution discussed above, but there are deep connections as we shall see shortly.

To illustrate the power and utility of (12), we first consider the asymptotic behaviour of $P(A,x)$ as $A \to \infty$ *without* making any assumption about the size of $x$. It is clear that in the limit $A \to \infty$ the integral is dominated by the contribution in the neighbourhood of $r = 0$, where $\tau_r \to \infty$. The following asymptotic expansions of $\mathrm{Ai}(z)$ and $\mathrm{Bi}(z)$ as $z \to \infty$ are therefore useful [19];

$$\text{Ai}(z) \sim \frac{1}{2}\pi^{-1/2} z^{-1/4} e^{-\frac{2}{3}z^{3/2}} \sum_{n=0}^{\infty} (-1)^n c_n z^{-3n/2}$$

$$\text{Bi}(z) \sim \pi^{-1/2} z^{-1/4} e^{\frac{2}{3}z^{3/2}} \sum_{n=0}^{\infty} c_n z^{-3n/2}$$

(13)

where

$$c_n = \frac{\Gamma(3n+\tfrac{1}{2})}{\Gamma(n+\tfrac{1}{2}) 36^n n!}.$$

(14)

Defining $f(z,x) \equiv \text{Ai}(z - xz^{-1/2})\text{Bi}(z) - \text{Bi}(z - xz^{-1/2})\text{Ai}(z)$, it is a straightforward although laborious task to show using (13) that as $z \to \infty$,

$$f(z,x) = \frac{1}{\pi z^{1/2}} \sinh x - \frac{1}{4\pi z^2}\left[x^2 \cosh x - x \sinh x\right] + O\!\left(\frac{1}{z^{7/2}}\right).$$

(15)

This is sufficient to develop the first two terms in an asymptotic expansion of $P(A,x)$ as $A \to \infty$ since, to the required order, (12) now simplifies to,

$$P(A,x) \sim \frac{e^x \sinh x}{\pi} \int_0^{\infty} \left[1 - \left(\frac{x^2 \coth x - x}{2} + \frac{5}{12}\right) r + O(r^2)\right] e^{-Ar - \frac{2}{3r}}\, dr.$$

(16)

One can easily evaluate the integrals in (16) using the result [20]

$$\int_0^{\infty} r^{\nu-1} e^{-\alpha r - \beta/r}\, dr = 2\left(\frac{\beta}{\alpha}\right)^{\nu/2} \hat{K}_\nu\!\left(2\sqrt{\alpha\beta}\right)$$

(17)

where $\hat{K}_\nu(z)$ is a modified Bessel function (we use $\wedge$ to distinguish $\hat{K}_\nu$ from $K_n$). The final step requires one to use the asymptotic expansion of $\hat{K}_\nu(z)$ as $z \to \infty$ [19]

$$\hat{K}_\nu(z) \sim \left(\frac{\pi}{2z}\right)^{1/2} e^{-z}\left\{1 + \frac{(4\nu^2-1)}{8z} + \frac{(4\nu^2-1)(4\nu^2-9)}{2!(8z)^2} + \ldots\right\} \quad (18)$$

whereupon one eventually derives as $A \to \infty$

$$P(A,x) \sim \frac{e^x \sinh x}{\sqrt{\pi}}\left(\frac{2}{3}\right)^{1/4}\frac{1}{A^{3/4}}\exp\left\{-\left(\frac{8}{3}\right)^{1/2}A^{1/2}\right\}$$

$$\times\left[1 - \frac{1}{A^{1/2}}\left(\frac{2}{3}\right)^{1/2}\left(\frac{13}{96} + \frac{x^2\coth x - x}{2}\right) + O\left(\frac{1}{A}\right)\right]. \quad (19)$$

We stress that this result is valid for *all* values of $x > 0$. For $x \to 0$ one recovers the leading order term conjectured in [15]. The relative size of the second term provides insight into the regime of validity of approximations made in the queueing problems discussed in [16,17]. With effort, one could calculate higher order terms in the asymptotic expansion if one felt so inclined.

For completeness, it is also possible to derive the behaviour of $P(A,x)$ as $A \to 0$. The exact result (12) is of little use in this regard, since the oscillatory nature of the integrand for large $r$ is difficult to handle. However, from the Laplace transform (6) we have as $s \to \infty$

$$\tilde{P}(s,x) = e^x \frac{\text{Ai}(2^{1/3}s^{1/3}x)}{\text{Ai}(0)}\left[1 - \frac{x^{1/2}}{2^{1/2}s^{1/2}} + O\left(\frac{1}{s^{2/3}}\right)\right]. \quad (20)$$

The leading order asymptotic behaviour of $P(A,x)$ as $A \to 0$ is governed by the first term in (20). This can be inverted exactly by noting that $\text{Ai}(z) = \pi^{-1}\sqrt{z/3}K_{1/3}(\tfrac{2}{3}z^{3/2})$ [19] and by appealing to (17). The result is

$$P(A,x) \sim \frac{2^{1/3}}{3^{2/3}\Gamma(\tfrac{1}{3})}\frac{xe^x}{A^{4/3}}e^{-2x^3/9A} \quad (21)$$

where we have used the fact that $\text{Ai}(0) = 3^{-2/3}/\Gamma(\frac{2}{3})$ [19]. In the limit $x \to 0$ this reduces to the exact distribution for the zero-drift case [15], valid for all $A > 0$. By reintroducing a general drift $\eta$ through the replacements $x \to \eta x$ and $A \to \eta^3 A$, one may readily see why this is the case.

Now we turn our attention to the moments. By considering the moments from two different perspectives we will show how to derive (3). We start by writing

$$\langle A^n \rangle \equiv \int_0^\infty A^n P(A,x)\, dA \tag{22}$$

into which we insert the formal expression (12) for the probability density $P(A,x)$. Interchanging the order of integration one has

$$\langle A^n \rangle = \frac{e^x n!}{\pi} \int_0^\infty \frac{\text{Ai}(\tau_r - x\tau_r^{-1/2})\text{Bi}(\tau_r) - \text{Bi}(\tau_r - x\tau_r^{-1/2})\text{Ai}(\tau_r)}{\text{Ai}^2(\tau_r) + \text{Bi}^2(\tau_r)} \frac{dr}{r^{n+1}}. \tag{23}$$

When $n = 0$ the right hand side of (23) is unity for all $x > 0$, although a direct demonstration of this fact is difficult and has so far eluded us. Notwithstanding, it follows that one cannot expand (23) in powers of $x$ when $n = 0$; mathematically the reason why may be traced backed to the observation that the integral

$$\int_0^\infty \frac{\text{Ai}(\tau_r)\text{Bi}(\tau_r)}{\text{Ai}^2(\tau_r) + \text{Bi}^2(\tau_r)} \frac{dr}{r^{n+1}} \tag{24}$$

is infinite when $n = 0$. However, for $n \geq 1$ this integral is finite and one can justify expanding the right hand side of (23) to at least $O(x^2)$. Using the fact that the Wronskian has a particularly simple form [19]

$$W\{\text{Ai}(z), \text{Bi}(z)\} \equiv \text{Ai}(z)\text{Bi}'(z) - \text{Ai}'(z)\text{Bi}(z) = \frac{1}{\pi} \tag{25}$$

one obtains after a change of variables

$$\langle A^n \rangle = \frac{xn!}{\pi^2} \int_0^\infty \frac{\tau_r^{-1/2}}{\mathrm{Ai}^2(\tau_r) + \mathrm{Bi}^2(\tau_r)} \frac{dr}{r^{n+1}} + O(x^2)$$

$$= \frac{3x}{4\pi^2} 2^{n+1} n! \int_0^\infty \frac{z^{3(n-1)/2}}{\mathrm{Ai}^2(z) + \mathrm{Bi}^2(z)} dz + O(x^2).$$

(26)

Next we consider the moments from the perspective of the Laplace transform, wherein $\langle A^n \rangle \equiv (-1)^n \partial_s^n \tilde{P}(s,x)\big|_{s=0}$. Expanding (6) as a power series in $x$ one has

$$\tilde{P}(s,x) = 1 + x + x 2^{1/3} s^{1/3} \frac{\mathrm{Ai}'(2^{-2/3} s^{-2/3})}{\mathrm{Ai}(2^{-2/3} s^{-2/3})} + O(x^2).$$

(27)

The function $\mathrm{Ai}'(z)/\mathrm{Ai}(z)$ has a well-known asymptotic expansion as $z \to \infty$ [1,12]

$$\frac{\mathrm{Ai}'(z)}{\mathrm{Ai}(z)} \sim 2 z^{1/2} \sum_{n=0}^\infty (-1)^n K_n z^{-3n/2}$$

(28)

where $K_n$ is defined by the quadratic recurrence relation (2). It is at this point that the connection with the Airy distribution becomes apparent. It follows that as $s \to 0$

$$\tilde{P}(s,x) \sim 1 + x \sum_{n=1}^\infty (-1)^n K_n 2^{n+1} s^n + O(x^2)$$

(29)

which implies that

$$\langle A^n \rangle = (-1)^n \left(\frac{\partial^n \tilde{P}}{\partial s^n}\right)\bigg|_{s=0} = x K_n 2^{n+1} n! + O(x^2).$$

(30)

If we now compare (30) with (26) and equate the coefficient of the $O(x)$ term we find immediately the result for $K_n$ given by (3). As a useful check one can evaluate (3) exactly when $n=1$. Thus, given the form of the Wronskian (25), one has that

$$\int_0^\infty \frac{dz}{\text{Ai}^2(z)+\text{Bi}^2(z)} = \pi\left[\tan^{-1}\frac{\text{Bi}(z)}{\text{Ai}(z)}\right]_0^\infty = \frac{\pi^2}{6} \tag{31}$$

where the final step uses the asymptotic forms for $\text{Ai}(z)$ and $\text{Bi}(z)$ given by (13) as well the fact that $\text{Ai}(0) = 3^{-2/3}/\Gamma(\tfrac{2}{3})$ and $\text{Bi}(0) = 3^{-1/6}/\Gamma(\tfrac{2}{3})$ [19]. It follows that $K_1 = 1/8$, as required. Numerical evaluation of (3) using Maple confirms its correctness for $2 \leq n \leq 10$. Further, when $n \to \infty$ one may use (13) to show that

$$K_n \sim \frac{3}{4\pi}\int_0^\infty z^{(3n-2)/2} e^{-\tfrac{4}{3}z^{3/2}}\, dz$$

$$= \frac{1}{2\pi}\left(\frac{3}{4}\right)^n \int_0^\infty y^{n-1} e^{-y}\, dy = \frac{1}{2\pi}\left(\frac{3}{4}\right)^n (n-1)!. \tag{32}$$

This result was first proved by Takacs [6,13] by studying directly the behaviour of (2) together with an alternative linear recursion relation for the $K_n$, namely

$$K_n = \frac{6n}{6n-1}c_n - \sum_{j=1}^{n-1} c_j K_{n-j} \tag{33}$$

where the coefficients $c_n$ are given by (14). The present derivation is much simpler, and can be extended to provide higher order terms almost trivially.

Having derived (3) somewhat indirectly, it is natural to ask whether there is a more direct derivation. The answer is yes, and we sketch the outline proof as follows (omitting the technical details). The starting point is (28), from which one can easily establish the two defining recursions for $K_n$, (2) and (33), using formal power series methods; see, e.g., [12]. With considerably more effort (the difficulties associated with (28) being an asymptotic expansion need careful handling) one can justify writing for $n \geq 1$

$$K_n = \frac{(-1)^n}{2\pi i} \int_C \frac{\text{Ai}'(z)}{\text{Ai}(z)} \frac{1}{2z^{1/2}} z^{3n/2} \frac{d(z^{3/2})}{z^{3/2}}$$

$$= \frac{(-1)^n}{2\pi i} \frac{3}{4} \int_C \frac{\text{Ai}'(z)}{\text{Ai}(z)} z^{3(n-1)/2} dz \qquad (34)$$

where the contour $C$ runs along two rays; the first from $\infty e^{-i2\pi/3}$ to 0 and the second from 0 to $\infty e^{i2\pi/3}$. Taking each ray in turn, making a straightforward change of variables, and noting that $e^{\pm i\pi(n-1)} = (-1)^{n-1}$, one can show that

$$K_n = \frac{1}{2\pi i} \frac{3}{4} \int_0^\infty \left[ e^{-i2\pi/3} \frac{\text{Ai}'(e^{-i2\pi/3} r)}{\text{Ai}(e^{-i2\pi/3} r)} - e^{i2\pi/3} \frac{\text{Ai}'(e^{i2\pi/3} r)}{\text{Ai}(e^{i2\pi/3} r)} \right] r^{3(n-1)/2} dr. \qquad (35)$$

It is now straightforward using (11) and (25) to show that (35) reduces to (3). It must be confessed, however, that the search for this second derivation was greatly assisted by knowing the answer first. This, coupled with the importance of (12) in its own right, explains why we have presented the analysis in the way that we have.

We conclude by making two observations which might prove worthy of further study. First, since $A_n = \langle A^n \rangle = (-1)^n \partial^n \tilde{P} / ds^n \big|_{s=0}$ it follows from (5) that

$$\frac{1}{2} \frac{d^2 A_n}{dx^2} - \frac{dA_n}{dx} = -nx A_{n-1} \qquad (36)$$

with the boundary condition $A_n(x=0) = 0$ for $n \geq 1$ and, by definition, $A_0(x) \equiv 1$. This differential equation has the formal solution

$$A_n(x) = 2n \int_0^x e^{2t} \int_t^\infty u e^{-2u} A_{n-1}(u) \, du \, dt \qquad (37)$$

which may be iterated to give any moment exactly, e.g., $A_1(x) = (x + x^2)/2$ etc. By taking the limit $x \to 0$ in (37) and by invoking (30) it follows that for $n \geq 1$

$$K_n = \frac{1}{2^n (n-1)!} \int_0^\infty u e^{-2u} A_{n-1}(u) \, du . \tag{38}$$

Whether such results in conjunction with (23) add something new is unclear. Second, consider a random variable $Y$ whose probability density is given by

$$f(y) = \frac{4}{\pi^2} \frac{y^{-1/3}}{\operatorname{Ai}^2(y^{2/3}) + \operatorname{Bi}^2(y^{2/3})} . \tag{39}$$

The moments of $Y$ are $Y_n \equiv 8 K_{n+1}$ for $n \geq 0$, which follows from (3) after a simple change of variables. Given the ubiquitous nature of $K_n$, it would be interesting to know whether this random variable $Y$ has a simple physical interpretation.

   In summary, we have succeeded in deriving an exact expression for the probability density for the area swept out by a drifted Brownian motion during its first passage time. Using this result it has proved possible to derive a simple closed-form solution for $K_n$, and hence the moments of the Airy distribution, in terms of the Airy functions $\operatorname{Ai}(z)$ and $\operatorname{Bi}(z)$. As well as being of mathematical interest, the results are applicable to a wide variety of physical problems.